\documentclass[prb,aps,preprint]{revtex4-1}
\usepackage{mathptmx}
\usepackage[T1]{fontenc}
\usepackage[utf8]{inputenc}
\usepackage{amsmath}
\usepackage{graphicx}

\makeatletter

\usepackage{dcolumn}
\usepackage{bm}
\usepackage{etoolbox}

\def\@email#1#2{%
 \endgroup
 \patchcmd{\titleblock@produce}
  {\frontmatter@RRAPformat}
  {\frontmatter@RRAPformat{\produce@RRAP{*#1\href{mailto:#2}{#2}}}\frontmatter@RRAPformat}
  {}{}
}%

\makeatother

\begin{document}
\title{Kinetic Cellular Model of Corrosion}
\author{Orkan Sezer}
\affiliation{Department of Earth Science and Engineering, Imperial College London,
South Kensington Campus, London, SW7 2AZ, United Kingdom}
\author{Andrew Horsfield}
\affiliation{Department of Materials and Thomas Young Centre, Imperial College
London, South Kensington Campus, London, SW7 2AZ, United Kingdom}
\email{a.horsfield@imperial.ac.uk}

\date{\today}
\begin{abstract}
Aqueous corrosion of metals involves multiple interconnected processes.
Thus, computer simulation of corrosion as a whole needs to be able
to describe the individual processes and how they influence each other.
Atomistic simulations are designed to obtain detailed information
for small regions of space over short times. Thus there are limits
to the understanding that can be obtained by atomistic simulations
alone. Here is presented a method that uses generalised rate equations
to extend the length and time scales that can be accessed. It is shown
to reduce to either the phase field or cellular automata methods in
certain limits. The generalised kinetic equations can reproduce the
behaviour described by both the Nernst-Planck and Butler-Volmer equations,
which are frequently used to describe corrosion. In addition, the
method can describe local rearrangements of atoms such as chemical
reactions. Example results are shown for illustrative 1D and 2D problems,
with good agreement being found with other methods. 
\end{abstract}
\maketitle

\section{Introduction}

Material degradation by corrosion significantly depreciates both the
monetary value and engineering reliability of the corroded substance.
Important infrastructure such as oil and gas pipelines, highways,
bridges, drinking water pipes, and sewer systems are examples of systems
that are especially susceptible to corrosion: this is due to the materials
used and their constant exposure to corrosive agents. The global annual
cost incurred from corrosion has been estimated to be 3.4\% of the
global Gross Domestic Product (GDP)\citep{koch2016} and has occasionally
culminated in loss of life.

The problems posed by anthropogenic climate change have motivated
many countries to invest in green technologies. Means to generate
clean energy include wave and tidal energy utilisation, as well as
off-shore wind power projects, due to the high energy densities of
waves and offshore winds\citep{vargas2019,uihlein2016}. Since such
methods of energy generation will have mechanical components either
fully or partially submerged in highly corrosive marine environments,
it is paramount not to just form a better understanding of corrosion
but also to develop computational methods that could serve to reliably
track corrosion propagation and inform lifetime analyses of corrodible
infrastructure components.

The simulation of corrosion is complex because of the number of processes
that need to be considered, and the range of length and time scales
over which they operate. Electron transfer from a metal to the neighbouring
solution can be very fast (fs). The dissolution and solvation of ions
are slower, but still take place at the atomic scale (nm) and are
relatively rapid ($\mu\mathrm{s}$). The diffusion and reactions of
these ions take place on longer time (ms) and length scales ($\mu\mathrm{m}$).
As there are charged particles involved, any model needs to consider
long range electrostatic interactions. Finally, as corrosion progresses,
the environment changes (e.g. corrosion product forms), modifying
the behaviour of the particles, and possibly introducing long ranged
elastic interactions. The influence of grain boundaries, multiple
phases, and microstructure more generally can also be important to
the above processes, making simulations still more complex.

Computational corrosion models have traditionally employed continuum
models of various kinds to manage the complexity of the problem. These
include rate equations\citep{de2016review, mccafferty1972kinetics, van1996formation, zhang2007mechanistic, nevsic2019implementation},
Finite Element Methods (FEMs)\citep{Xu2013,Liu2019}, Cellular Automata
(CA)\citep{khatami2021,jafarzadeh2019,di20163d,cui2019experimental,guiso2020intergranular,cordoba2001cellular,di2011morphology,chen2017simulation,malki2005computer},
a hybrid of CA and FEM \citep{fatoba2018,wang2016computational},
and Phase Field (PF) \citep{cui2021,Guyer2004,Guyer2004a,ansari2018phase,ansari2020multi,bischoff2021band,chadwick2018numerical,cui2022generalised,gao2020efficient,gao2020space,hu2022meso,lin2019phase,lin2020multi,mai2016phase,mai2017phase,mai2018new,martinez2018phase,nguyen2017modeling,nguyen2017phase,nguyen2018phase,wen2012phase,xiao2018quantitative}.
FEMs aim to solve the Nernst-Planck equation with suitable boundary
conditions (such as the Butler-Volmer equation) but require additional
algorithmic steps to link geometry and phase. CA, on the other hand,
have been popular as they can reflect the stochastic nature of corrosion
and are able to represent multi-phase and multi-species systems without
great complexity. PF models have been widely used to solve interfacial
problems of multi-phase systems \citep{steinbach2013}, and in corrosion
models, PF aims to solve equations that describe the temporal evolution
of both phase and concentration \citep{cui2021,Guyer2004,Guyer2004a}.

However, it is increasingly realised that atomistic processes (notably
at interfaces) are important. There have been a number of studies
using Density Functional Theory (DFT) of corrosion and of electrochemical
problems more generally \citep{DARBY2022101118,GRO2023101170,GRO2023101345,tang2020modeling}.
They have generated insights, and the techniques are improving rapidly.
To improve the description of the extended environment, hybrid schemes
have been developed that combine DFT with atomistic potentials, and
dielectric continua \citep{lischner2011joint,Yang2022recent}. These
extensions are important, but the time scales at least are set by
the DFT part of the simulation. Going further than this has tended
to mean returning to PF or other continuum models.

Here is presented a model that captures atomic scale information,
while allowing extended length and time scales to be addressed. The
approach taken is to use generalised kinetic equations. The Kinetic
Cellular Model (KCM) views a system as composed of cells that exchange
energy and particles with each other following kinetic equations,
while allowing chemical reactions to take place within the cells.
The method thus allows modelling of diffusive processes and chemical
reactions, and adds electrostatics once charged species are involved.
The KCM is thus close to a multi-field universal corrosion model theorized
by Jafarzadeh \textit{et al} \citep{jafarzadeh2019}. While KCM can
include chemical reactions, the theory is still at an early stage.
Mechanical stress fields have not been introduced yet, though it is
possible to see in principle how this might be done by giving the
material in each cell a strain state. To keep the number of additional
parameters low, chemical reactions are currently assumed to be non-catalytic,
non-inhibited, and do not change the temperature. We believe that
lifting these constraints should be possible. Looking to the future,
we hope that the structure of this method should allow detailed models
of free energies based on DFT calculations to be introduced, including
a more accurate treatment of interfacial energies than is usually
the case in the PF method. Finally, for electrochemical processes,
electrons need to be treated explicitly: a simple scheme in which
electrons are characterized by a charge and a chemical potential has
already been introduced, but this can be improved. We note that this
method has similarities with that of Watanabe and Fujita \citep{Watanabe2022}.

The KCM software developed here is applied to simple model problems.
The results and current limitations can be found in the Results section.
Future work is discussed in the Conclusions section.

\section{Methodology}

As discussed above, corrosion involves coupled processes that span
multiple length and time scales. To simulate corrosion we thus need
a method that embraces all the important processes at their respective
length and time scales. This is a difficult task. The solution proposed
here is to divide a system into cells and permit material to be transported
between neighbouring cells. This creates a way to identify interfaces
and, if the rule governing the flow of particles is sufficiently general,
to simulate both diffusion of mobile species and activated events
at interfaces, such as the dissolution of metal ions.

In the KCM there are two types of process: the diffusion of particles
and the local rearrangement of particles. The same division is used
in the PF method with conservative fields (diffusion) and non-conservative
fields (phase change)\citep{qin2010}. The fundamental physical assumption
is that cells are (nearly) in equilibrium with themselves, but not
necessarily in equilibrium with their neighbours. The internal rearrangement
is clearly a result of a lack of internal equilibrium, but how this
is handled will depend on the rate of that rearrangement relative
to the rate of transfer of particles between cells. If internal rearrangement
is very slow, then we can treat the cell as being in some quasi-equilibrium
state for the purposes of computing particle transfer. If it is very
fast, then the rearrangement will happen essentially instantaneously
on the time scale of particle transfer. In this case the final arrangement
can be computed simply as a function of particle concentration, meaning
the cell is permanently in equilibrium from the viewpoint of particle
transfer. If the diffusion of particles between cells and the rearrangement
of the particles within cells occur with similar rates, the method
will be less accurate.

In what follows we will describe briefly the key ideas of the KCM,
and establish a link between the KCM and a combination of the Nernst-Planck
and Butler-Volmer equations: a combination used in finite element
calculations of corrosion \citep{Xu2013}. We will also note the link
with the PF and CA methods.

\subsubsection{Particle Transfer Between Cells}

We treat each cell, labelled by $i$, as being locally in equilibrium,
and then associate a number of particles $n_{\alpha,i}$ and an electrochemical
potential $\mu_{\alpha,i}$ for each species, labelled by $\alpha$.
We then imagine random hopping of particles between cells described
by the following rate equation 
\begin{equation}
\frac{\partial n_{\alpha,i}}{\partial t}=\sum_{j}\left[n_{\alpha,j}R_{\alpha,j\to i}-n_{\alpha,i}R_{\alpha,i\to j}\right]\label{eq:CPF-01}
\end{equation}
where $R_{\alpha,i\to j}$ is the rate at which one particle of type
$\alpha$ hops from cell $i$ to nearest neighbor cell $j$. The energy
change per species is given by the number of particles of each species
exchanged times the difference in electrochemical potential between
the cells for that species. Note that there must not be a configurational
entropy term in the chemical potential (apart possibly from a term
purely internal to the cell) as this is taken care of by the hopping
between cells.

This approach has similarities with Lattice Kinetic Monte Carlo (LKMC)\citep{andersen2019practical}
and Master Equations (ME)\citep{gillespie1992rigorous}, but differs
from both. Unlike LKMC all particles evolve at the same time rather
than considering individual hops of particles. Unlike ME, it is concentrations
that evolve with time rather than probabilities of configurations,
which makes it a set of coupled rate equation.

One way to estimate the rate $R_{\alpha,i\to j}$ that gives the correct
equilibrium distribution is to use 
\begin{equation}
R_{\alpha,i\to j}=\nu_{\alpha}\exp\left(-\frac{\mu_{\alpha,j}-\mu_{\alpha,i}}{2k_{B}T}\right)\label{eq:CPF-02}
\end{equation}
where $\nu_{\alpha}$ is the attempt frequency for species $\alpha$,
$k_{B}$ is Boltzmann's constant, and $T$ is the temperature. The
electrochemical potential is given by $\mu_{\alpha}=\bar{\mu}_{\alpha}+q_{\alpha}v$,
where $\bar{\mu}_{\alpha}$ is the chemical potential, $q_{\alpha}$
is the charge of species $\alpha$, and $v$ is the electrostatic
potential. Note that this rule is closely related to the Butler-Volmer
equation. The rate equation given in Eq. \ref{eq:CPF-01}, together
with the expression for the rate given in Eq. \ref{eq:CPF-02}, form
the core of that part of the KCM associated with particle transfer.

The electrostatic potential $v$ is found from Poisson's equation,
which takes the following form in the presence of dielectric screening
\begin{equation}
\vec{\nabla}\cdot\left(\epsilon_{0}\epsilon_{r}\vec{\nabla}v\right)=-\rho\label{eq:CPF-02a}
\end{equation}
where $\epsilon_{r}$ is the relative permittivity of the medium,
$\rho$ is the charge density, and $\epsilon_{0}$ is the permittivity
of free space. The charge density in cell $i$ is given by 
\begin{equation}
\rho_{i}=\sum_{\alpha}ez_{\alpha}\frac{n_{\alpha,i}}{\Omega}\label{eq:CPF-02b}
\end{equation}
where $e$ is the magnitude of the charge on an electron, $z_{\alpha}$
is the charge (in units of $e$) of species $\alpha$, and $\Omega$
is the volume of a cell. The charge density is assumed to be uniform
within a cell.

To illustrate how the KCM works, consider a system that is one dimensional.
If the hopping occurs only between nearest neighbour cells separated
by a distance $a$ then the net change in the number $\tilde{n}_{\alpha,i}$
of particles per unit area of type $\alpha$ in cell $i$ after a
short time $\delta t$ is 
\begin{align}
\delta\tilde{n}_{\alpha,i} & =\delta t\left(\tilde{n}_{\alpha,i+1}R_{\alpha,i+1\to i}-\tilde{n}_{\alpha,i}R_{\alpha,i\to i+1}\right)\nonumber \\
 & +\delta t\left(\tilde{n}_{\alpha,i-1}R_{\alpha,i-1\to i}-\tilde{n}_{\alpha,i}R_{\alpha,i\to i-1}\right)\label{eq:CPF-03}
\end{align}
Eq. \ref{eq:CPF-03} can be viewed as a cellular automata update rule.
Thus we see that CA is a limiting case where the cell concentrations
have a restricted set of allowed values. If the difference in chemical
potential between neighbouring cells is small, then from Eq. \ref{eq:CPF-02}
we have \citep{lin2019phase} 
\begin{equation}
R_{\alpha,i\to j}\approx\nu_{\alpha}\left[1-\frac{\mu_{\alpha,j}-\mu_{\alpha,i}}{2k_{B}T}\right].\label{eq:CPF-04}
\end{equation}
It is then straightforward to derive the Nernst-Planck equation in
one dimension, which can then be generalized to three dimensions.
Note that, to reproduce the Nernst-Planck equation, we require that
$\nu_{\alpha,i}=\frac{1}{a^{2}}D_{\alpha,i}$, where $D_{\alpha,i}$
is the diffusivity for species $\alpha$ in cell $i$. We can also
write $\nu_{\alpha,i}$ in terms of the mobility $m_{\alpha,i}$:
$\nu_{\alpha,i}=\frac{k_{B}T}{a^{2}e}m_{\alpha,i}$. We note that
this limit is also closely related to the PF equation of motion for
a conserved field\citep{qin2010}. Thus we see that a fundamental
assumption of the PF equation of motion is the slow variation of electrochemical
potential with position.

\subsubsection{Particle Rearrangement Within Cells}

We now consider processes that occur within a cell. As noted above,
if the rearrangements are fast they can be treated as instantaneous.
One example of a fast chemical reaction is 
\begin{equation}
e^{-}+H^{+}\rightleftharpoons\frac{1}{2}H_{2}.\label{eq:CPF-04a}
\end{equation}
This reaction removes an electron and a proton and creates half a
hydrogen molecule. In this case we just need to know the amount of
each species present before the reaction, from which we compute the
amount of each product species after the reaction, and no rate equations
are needed.

By contrast, changes of phase in solids can be slower because they
are themselves limited by diffusion. In this case a rate equation
within a cell will be needed. The exact form of the equation will
depend on the process. For some cases we can probably borrow the results
from the PF method\citep{qin2010}.

Note that the derivative of the free energy $\Delta G_{F}$ with respect
to the number of particles $n_{\alpha,i}$ is just the electrochemical
potential $\mu_{\alpha,i}$ 
\begin{equation}
\mu_{\alpha,i}=\frac{\partial\Delta G_{F}}{\partial n_{\alpha,i}}\label{eq:CPF-05}
\end{equation}
which allows us to connect the local rearrangements within a cell
(given by the rate of change of free energy with phase) to the rate
of particle transfer between cells in a consistent manner. In addition,
the chemical potential in Eq. \ref{eq:CPF-05} can be related to the
PF method through the functional derivative of the free energy with
number density $c_{\alpha}$ of particles of type $\alpha$ by the
following expression 
\begin{equation}
\mu_{\alpha,i}=\frac{\partial\Delta G_{F}}{\partial n_{\alpha,i}}=\left\langle \frac{\delta\Delta G_{F}}{\delta c_{\alpha}}\right\rangle _{i},\label{eq:CPF-05a}
\end{equation}
where the average value is taken by integrating over the volume of
cell $i$ and then dividing by that volume.

\section{Results and discussion}

As a test of the method, four simple one dimensional problems are
considered: the diffusion of particles, the Gouy-Chapman distribution
of ions in solution, the evolution of hydrogen from a cathode, and
the dissolution of Mg into water. We conclude with a discussion of
simulations of diffusion in 2D and 3D.

\subsection{Diffusion}

\begin{figure}
\centering{}\includegraphics[width=1\columnwidth]{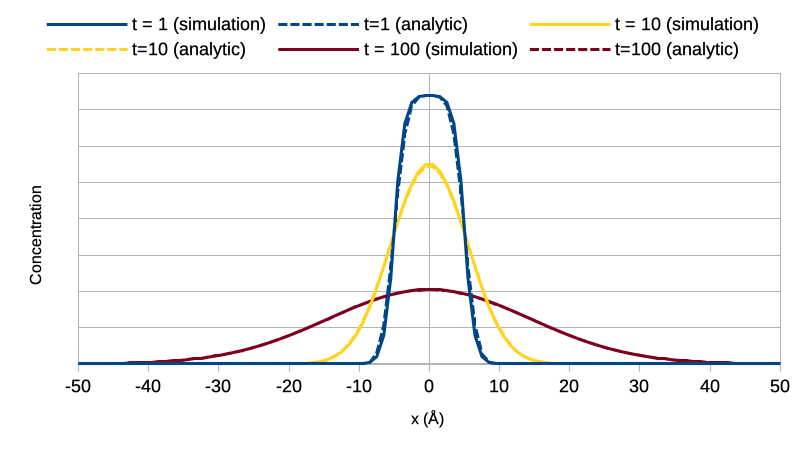} \caption{ This figure shows the diffusion profiles at three different times.
Both the solutions from the KCM method (solid lines), and the analytic
solution (dashed lines) are shown. The time is in units of Å$^{2}/D$,
where $D$ is the diffusion coefficient, and the concentration is
in arbitrary units. Note that the two sets of solutions are almost
indistinguishable.}\label{fig:diffusion}
\end{figure}

The first, and most simple, problem is the diffusion of particles
with diffusion coefficient $D$ that begin in a narrow region of width
$w$ with uniform concentration $c_{0}$ per unit volume. This problem
can be solved analytically, and the concentration of particles at
time $t$ and position $x$ is given by 
\begin{equation}
c(x,t)=\frac{c_{0}}{2}\left\{ \mathrm{erf}\left(\frac{x}{\sqrt{4Dt}}\right)-\mathrm{erf}\left(\frac{x-w}{\sqrt{4Dt}}\right)\right\} .\label{eq:CPF-06}
\end{equation}
From Fig. \ref{fig:diffusion} we see we get essentially perfect agreement
between the analytic solution and the results of the KCM simulation.

\subsection{Gouy-Chapman distribution}

The second problem is to find the distribution of ions in a solution
containing equal amounts of positive and negative ions (such as NaCl
dissolved in water) between two plates of equal and opposite charge.
Again, there is an analytic solution. There is also a well known linear
approximation to it that assumes small changes in the potential \citep{Gouy1910,Chapman1913},
which is what we use here as it is sufficiently accurate for our purposes.
At equilibrium $\frac{\partial n_{\alpha,i}}{\partial t}=0$, and
hence, from Eqs. \ref{eq:CPF-01} and \ref{eq:CPF-02} we have 
\begin{equation}
n_{\alpha,i}\exp\left(\frac{\mu_{\alpha,i}}{k_{B}T}\right)=n_{\alpha,j}\exp\left(\frac{\mu_{\alpha,j}}{k_{B}T}\right)\label{eq:CPF-06a}
\end{equation}
which has the solution 
\begin{equation}
n_{\alpha,i}=\bar{n}_{\alpha}\exp\left(-\frac{\mu_{\alpha,i}-\bar{\mu}_{\alpha}}{k_{B}T}\right)\label{eq:CPF-06b}
\end{equation}
where $\bar{n}_{\alpha}$ and $\bar{\mu}_{\alpha}$ are reference
populations and electrochemical potentials and are set to ensure the
total number of particles of type $\alpha$, $N_{\alpha}$, is correct:
$\sum_{i}n_{\alpha,i}=N_{\alpha}$. We also have $\mu_{\alpha,i}=\bar{\mu}_{\alpha,i}+z_{\alpha}ev_{i}$,
where $\bar{\mu}_{\alpha,i}$ is the chemical potential of species
$\alpha$ in cell $i$, $z_{\alpha}$ is the charge on species $\alpha$,
and $v_{i}$ is the potential in cell $i$. We can then rewrite Eq.
\ref{eq:CPF-06c} as

\begin{equation}
n_{\alpha,i}=\tilde{n}_{\alpha,i}\exp\left(-\frac{z_{\alpha}ev_{i}}{k_{B}T}\right)\label{eq:CPF-06c}
\end{equation}
where 
\begin{equation}
\tilde{n}_{\alpha,i}=\bar{n}_{\alpha}\exp\left(-\frac{\bar{\mu}_{\alpha,i}-\bar{\mu}_{\alpha}}{k_{B}T}\right)\label{eq:CPF-06d}
\end{equation}
and $\tilde{n}_{\alpha,i}$ is the population in regions of zero potential.
For the ions, $\bar{\mu}_{\alpha,i}$ can be assumed to be independent
of position (dilute solution), and hence $\tilde{n}_{\alpha,i}$ makes
Eq. \ref{eq:CPF-06c} the standard Poisson-Boltzmann result. Provided
the potential variation is small, we can derive an analytic expression
for the potential $v(x)$ using Gouy-Chapman theory \citep{Gouy1910,Chapman1913}
\begin{equation}
v(x)=\frac{\Delta v_{0}}{2}\frac{\sinh(\kappa x)}{\sinh(\kappa l)}\label{eq:CPF-07}
\end{equation}
where $\Delta v_{0}$ is the potential drop between the two electrodes,
which are separated by a distance of $2l$, and 
\begin{equation}
\kappa=\sqrt{\frac{\sum_{\alpha}(z_{\alpha}e)^{2}\bar{n}_{\alpha}}{\epsilon_{0}\epsilon_{r}k_{B}T\Omega}}\label{eq:CPF-08}
\end{equation}
is the inverse Debye screening length. Note that the ions are forbidden
to enter the metal electrodes and the electrons are forbidden to enter
the solution: they are each given a diffusion constant of zero in
their respective forbidden regions. The potential difference is established
by giving the electrons different chemical potentials in the two electrodes:
the difference in this simulation is 0.1 eV. This chemical potential
difference results in electrons transferring from the right electrode
(which acquires a positive charge) to the left electrode (which acquires
a negative charge).

As we seek an equilibrium configuration, the steady state solution
was sought directly rather than rate equations being solved. This
was achieved by a steepest descent minimization of an effective free
energy $G$. The free energy has a form chosen such that its minimum
gives the correct steady state particle distribution (Eq. \ref{eq:CPF-06c}),
namely 
\begin{equation}
G=\sum_{\alpha i}n_{\alpha,i}\left\{ k_{B}T\left[\ln\left(\frac{n_{\alpha,i}}{\tilde{n}_{\alpha,i}}\right)-1\right]+\frac{1}{2}z_{\alpha}ev_{i}\right\} +\sum_{\alpha}\bar{\mu}_{\alpha}N_{\alpha}.\label{eq:CPF-09}
\end{equation}
If we substitute Eq. \ref{eq:CPF-06d} into Eq. \ref{eq:CPF-09} we
obtain 
\begin{align}
G & =\sum_{\alpha i}n_{\alpha,i}\left\{ k_{B}T\left[\ln\left(\frac{n_{\alpha,i}}{\bar{n}_{\alpha}}\right)-1\right]+\bar{\mu}_{\alpha,i}+\frac{1}{2}z_{\alpha}ev_{i}\right\} \nonumber \\
 & +\sum_{\alpha}\bar{\mu}_{\alpha}\left(N_{\alpha}-\sum_{i}n_{\alpha,i}\right)\label{eq:CPF-09a}
\end{align}
which is more intuitive. We now see that $\bar{\mu}_{\alpha}$ corresponds
to a Lagrange multiplier enforcing the conservation of number of particles
of type $\alpha$. To find the minimum of $G$ we differentiate it
with respect to $n_{\alpha,i}$ and set the derivatives to zero. This
gives 
\begin{equation}
\frac{\partial G}{\partial n_{\alpha,i}}=k_{B}T\ln\left(\frac{n_{\alpha,i}}{\bar{n}_{\alpha}}\right)+\bar{\mu}_{\alpha,i}-\bar{\mu}_{\alpha}+z_{\alpha}ev_{i}=0\label{eq:CPF-10b}
\end{equation}
which has Eq. \ref{eq:CPF-06b} as its solution. Provided we can neglect
$\frac{\partial v_{i}}{\partial n_{i}}$, we can use this result to
produce the following Newton-Raphson formula for updating the numbers
of particles of species $\alpha$ 
\begin{equation}
n^{(k+1)}_{\alpha,i}=n^{(k)}_{\alpha,i}\left[1+f\ln\left(\frac{n^{(k)}_{\alpha,i,out}}{n^{(k)}_{\alpha,i}}\right)\right]\label{eq:CPF-10c}
\end{equation}
where $k$ is the iteration number in the minimization, $1\ge f>0$
is a damping factor to ensure smooth convergence, and $n_{\alpha,i,out}$
is computed from Eq. \ref{eq:CPF-06b}.

\begin{figure}
\centering{}\includegraphics[width=1\columnwidth]{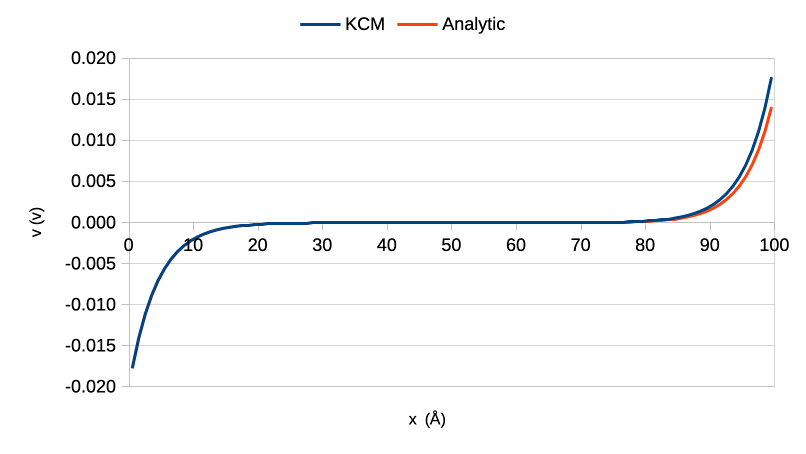}\caption{ The variation of potential with position in a solution of NaCl in
water between two charged electrodes. Note the close agreement between
the numerical result found using KCM (blue line), and the analytic
result found using Gouy-Chapman theory (red line). See the main text
for an explanation of the small differences.}\label{fig:ions-potential}
\end{figure}

Using the parameters for the simulation ($\left|z_{\alpha}\right|=1$,
$\bar{n}_{\alpha}=0.01$, $\epsilon_{r}=100$, $T=300\,$K, and $\Omega=27\,$Å$^{3}$)
we have $\kappa=0.23\,$Å$^{-1}$. From Fig. \ref{fig:ions-potential}
we see that there is good agreement between the Gouy-Chapman result
and the potential found using the KCM method. Small differences are
expected as the KCM method does not make the linear approximation
assumed by the Gouy-Chapman theory, and the electrodes are simulated
as metals with electrons that vary their density over a finite distance,
and which can respond to fields generated by the ions in solution,
rather than thin sheets of fixed charge.

\subsection{Hydrogen evolution}

The third problem, which is designed to show the capabilities of the
method and may not be especially realistic, is the reaction of excess
protons in solution with excess electrons in a metal electrode (cathode)
to form hydrogen gas. This is meant to represent the hydrogen evolution
reaction that appears at a cathode in corrosion. The system consists
of three parts: the metal electrode containing excess electrons, a
film of corrosion product on top of the electrode, and a solution
containing excess protons. The parameters are set so that protons
cannot diffuse into the metal (diffusion coefficient of zero), but
they can diffuse in the film and solution. The electrons can diffuse
anywhere, but diffuse 37.5 times more quickly in the metal than elsewhere.
The chemical potentials are arranged to steer particles in the expected
directions: the electrons have a chemical potential of 0 eV in the
metal, 1.0 eV in the film, and 5.0 eV in the solution; the protons
have a chemical potential of 0 eV in the metal and -1.0 eV in both
the film and the solution. The dielectric function is set to 100 in
the metal, 10 in the film and 80 in the solution. The simulation is
carried out at room temperature.

\begin{figure}
\centering{} \includegraphics[width=0.75\columnwidth]{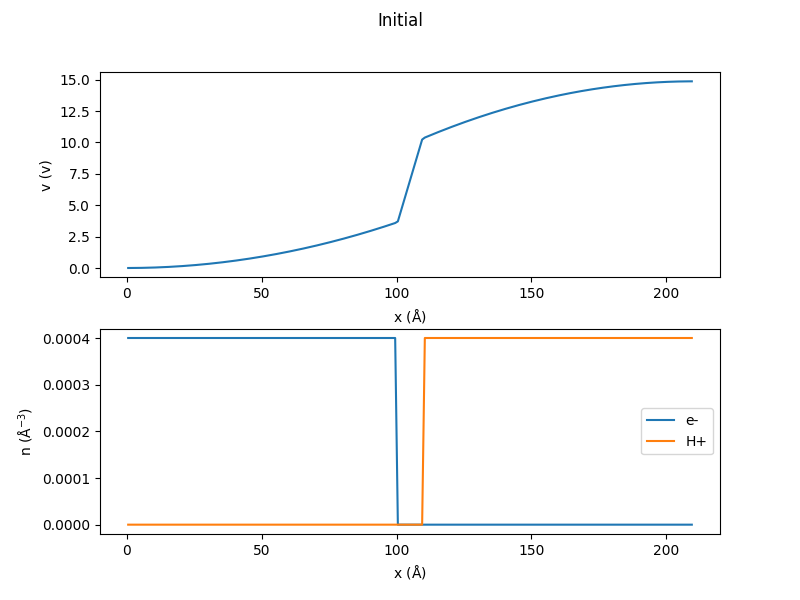}\\
 \includegraphics[width=0.75\columnwidth]{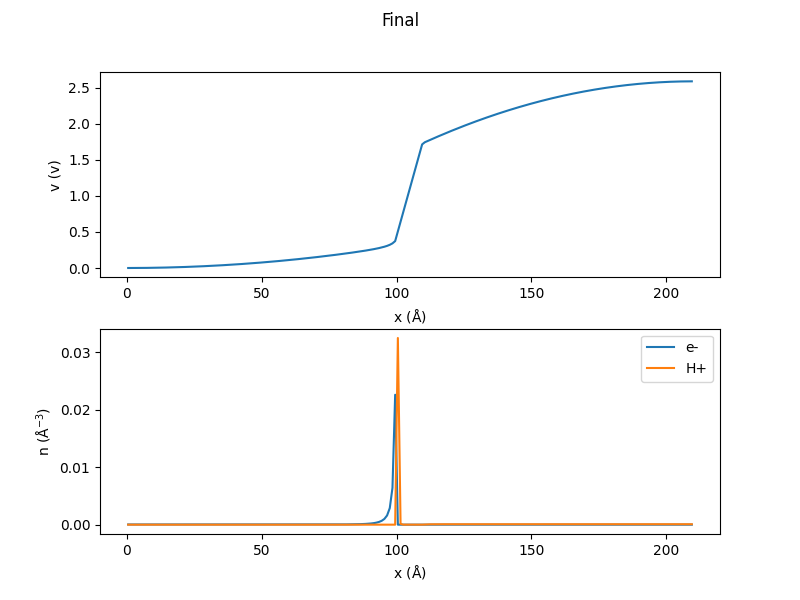} \caption{ The initial and final distributions of the electrons and protons
with the associated potentials. The region below $x=100\,$Å\ is
the metal, and is where the excess electrons begin. The region above
$x=110\,$Å\ is the water in which the excess protons begin. Between
them is the film of corrosion product. The potential has a much steeper
slope in the film region than elsewhere because of the lower dielectric
constant (10 in the film, 100 in the metal, 80 in the water).}\label{fig:electrons-protons}
\end{figure}

The initial state of the simulation is shown in the upper panel of
Fig. \ref{fig:electrons-protons}. Very quickly the electrons and
protons diffuse to the metal/film interface, attracted by the electrostatic
interaction between them. The electrons then slowly diffuse out of
the metal to react with the protons. As the reaction of the electrons
with the protons is instantaneous, the rate limiting step is the transfer
of electrons from the metal to the film. The resulting distribution
after partial reaction of the protons with the electrons to form hydrogen
gas is shown in the lower panel of Fig. \ref{fig:electrons-protons}.
The potential difference between the left and the right of the system
is produced by the dipole layer at the interface formed by the remaining
protons and electrons.

\subsection{Dissolution of solid into solution}

To simulate corrosion we need to include the dissolution of solid
metal into solution as ions. The ions can then react with the water
to form solid corrosion product: we will not consider this additional
step at this point.

\begin{figure}
\centering{}\includegraphics[width=1\columnwidth]{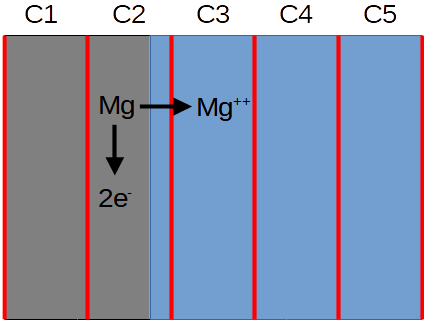} \caption{This is a representation of the $\mathrm{Mg/H_{2}O}$ interface.
The grey region to the left is Mg metal, and the blue region to the
right is water. The vertical red lines indicate the boundaries of
the cells used in the algorithm. Most cells contain just one phase.
However, cell C2 contains a mixture of metal and water. The formation
of these mixed cells is a consequence of metal atoms being transferred
from the solid metal into solution, creating space that can then be
filled with water.}\label{fig:Mg_H2O}
\end{figure}

Dissolution requires us to make a fundamental modification to the
way the algorithm operates. Until now we have assumed each cell contains
only one phase, while dissolution requires that any given cell be
able to contain more than one. Consider the corrosion of Mg (see Fig.
\ref{fig:Mg_H2O}). When Mg atoms from solid Mg metal dissolve in
the water to form $\mathrm{Mg}^{++}$ ions, space is created in the
metal that can now be filled with water. When this happens we create
a cell that has a mixture of phases.

The transfer of atoms between cells continues to be determined by
the chemical potentials of the species within them. If a cell is completely
filled with solid metal, then water is excluded, and thus the chemical
potential for water in that cell is large and positive. Once space
becomes available, a limited amount of water can be added, and the
corresponding chemical potential decreases. Once water has filled
the vacant space it is difficult again for further water to be added,
and the chemical potential increases again. Thus we see that chemical
potentials now need to have a term that accounts for the change in
phase. The chemical potential of $\mathrm{Mg}^{++}$ ions in water
is given by the concentration of ions, with the saturation being marked
by a rise in chemical potential.

We now need to consider what happens within cells. We treat the contents
of the cell as being a kind of average phase. While this is an approximation,
treating the phases separately is equivalent to dividing a cell into
subcells with new interfaces. The natural way to handle this would
be to start with smaller cells. However, we still need to ensure there
is as much consistency as possible between reactions within cells
and the transfer of atoms from one cell to another: this can be achieved
by a consistent model for the free energy from which chemical potentials
are then derived.

Here we will represent the average local environment through a phase
field parameter $\phi$ which satisfies 
\begin{equation}
\phi_{i}=f\left(\frac{n^{(\mathrm{effective})}_{\mathrm{Mg},i}+n^{(\mathrm{effective})}_{\mathrm{Mg++},i}}{n^{(\mathrm{solid})}_{\mathrm{Mg}}},w\right)\label{eq:CPF-12}
\end{equation}
where $n^{(\mathrm{effective})}_{\mathrm{Mg},i}$ is an effective
number of Mg atoms for cell $i$ that incorporates information from
the neighbourhood of the cell. This allows a surface to be distinguished
from the bulk, for example. We will use

\begin{align}
n^{(\mathrm{effective})}_{\mathrm{Mg},i} & =\frac{1}{7}\left(n_{\mathrm{Mg},i-1}+5n_{\mathrm{Mg},i}+n_{\mathrm{Mg},i+1}\right)\label{eq:CPF-12a}\\
n^{(\mathrm{effective})}_{\mathrm{Mg++},i} & =\frac{1}{7}\left(n_{\mathrm{Mg++},i-1}+5n_{\mathrm{Mg++},i}+n_{\mathrm{Mg++},i+1}\right)
\end{align}
where $n_{\mathrm{Mg},i}$ is the number of Mg atoms in cell $i$,
$n_{\mathrm{Mg++},i}$ is the number of $\mathrm{Mg}^{++}$ ions in
cell $i$, and $n^{(\mathrm{solid})}_{\mathrm{Mg}}$ is the corresponding
value for bulk solid Mg metal. The weighting factors are based on
the idea that space is divided into cubes, and each cube has six neighbouring
cubes. The effective number of Mg or $\mathrm{Mg}^{++}$ atoms is
then just the average of the numbers for each of the seven cubes constructed
from the central cube and its six neighbours. As the model is 1 dimensional
the four neighbouring cubes in the same plane as the central cube
will have the same number of atoms as the central cube, hence the
factor of 5. The width $w$ sets how rapidly the phase changes with
number of Mg atoms, and is set to 0.01 in these calculations. We will
assume the function $f$ has the form 
\begin{equation}
f(x_{i},w)=\begin{cases}
0 & x_{i}<1-3w\\
\frac{x_{i}-1+3w}{2w} & 1-3w\le x_{i}\le1-w\\
1 & x_{i}>1-w
\end{cases}\label{eq:CPF-13}
\end{equation}
where 
\begin{equation}
x_{i}=\frac{n^{(\mathrm{effective})}_{\mathrm{Mg},i}+n^{(\mathrm{effective})}_{\mathrm{Mg++},i}}{n^{(\mathrm{solid})}_{\mathrm{Mg}}},\label{eq:13a}
\end{equation}
$f=0$ corresponds to the aqueous region, and $f=1$ corresponds to
the solid metal region.

When the Mg dissolves it transforms into aqueous $\mathrm{Mg}^{++}$
ions plus some free electrons according to 
\begin{equation}
\mathrm{Mg}\rightleftharpoons\mathrm{Mg}^{++}+2e^{-}\label{eq:13b}
\end{equation}
We treat this as a spontaneous reaction that takes place when Mg is
in direct contact with water. This is determined by looking at the
phase field parameter: if this is less than 0.95 the Mg is considered
to be in contact with water.

To obtain a model for our chemical potentials, we write down a free
energy expression for our system and then use Eq. \ref{eq:CPF-05}.
For the case of Mg dissolving in water, our free energy is 
\begin{align}
G & =\sum_{i}\left(n_{\mathrm{Mg},i}g_{\mathrm{Mg},i}+n_{\mathrm{Mg}++,i}g_{\mathrm{Mg}++,i}+n_{\mathrm{H_{2}O},i}g_{\mathrm{H_{2}O},i}\right)\nonumber \\
 & +\sum_{i}\Delta n_{\mathrm{e},i}g_{\mathrm{e},i}+\frac{1}{2}\sum_{i}\left(2n_{\mathrm{Mg}++,i}-\Delta n_{e,i}\right)ev_{i}\label{eq:CPF-10}
\end{align}
where, for cell $i$, $g_{\mathrm{Mg},i}$ is the free energy per
particle for Mg, $g_{\mathrm{Mg}++,i}$ is the free energy per particle
for $\mathrm{Mg}^{++}$, $n_{\mathrm{H_{2}O},i}$ is the number of
water molecules, $g_{\mathrm{H_{2}O},i}$ is the free energy per molecule
for water, $\Delta n_{\mathrm{e},i}$ is the number of free electrons,
and $g_{\mathrm{e},i}$ is the free energy per free electron. The
free energies per particle exclude the long ranged electrostatic interactions,
which we treat explicitly through the electrostatic potential $v_{i}$.
The charge density in cell $i$ is given by $\rho_{i}=e\left(2n_{\mathrm{Mg}++,i}-\Delta n_{e,i}\right)/\Omega$
and is assumed to be uniform within a cell. The results presented
here are for simulations in 1D only, so Poisson's equation is solved
in 1D. Note that this is very similar to Eq. \ref{eq:CPF-09a}, but
without the configurational entropy or particle number conserving
terms: these are accounted for by the equations of motion (for example,
see Eq. \ref{eq:CPF-06b}).

We will assume that the water does not change much during the dissolution
of Mg so that the corresponding free energy $G_{\mathrm{H_{2}O}}=\sum_{i}n_{\mathrm{H_{2}O},i}g_{\mathrm{H_{2}O},i}$
can be treated as constant. The free energy we will work with is then
\begin{align}
\Delta G & =G-G_{\mathrm{H_{2}O}}\nonumber \\
 & =\sum_{i}\left(n_{\mathrm{Mg},i}g_{\mathrm{Mg},i}+n_{\mathrm{Mg}++,i}g_{\mathrm{Mg}++,i}\right)\nonumber \\
 & +\sum_{i}\Delta n_{\mathrm{e},i}g_{\mathrm{e},i}+\frac{1}{2}\sum_{i}\left(2n_{\mathrm{Mg}++,i}-\Delta n_{e,i}\right)ev_{i}\label{eq:CPF-11}
\end{align}
We will use the following linear interpolation formulae to introduce
the phase dependence of the free energy per particle for electrons
and Mg atoms 
\begin{align}
g_{\mathrm{e},i} & =\phi_{i}g^{(\mathrm{solid})}_{\mathrm{e},i}+\left(1-\phi_{i}\right)g^{(\mathrm{water})}_{\mathrm{e},i}\nonumber \\
g_{\mathrm{Mg}++,i} & =\phi_{i}g^{(\mathrm{solid})}_{\mathrm{Mg}++,i}+\left(1-\phi_{i}\right)g^{(\mathrm{water})}_{\mathrm{Mg}++,i}\\
g_{\mathrm{Mg},i} & =\phi_{i}g^{(\mathrm{solid})}_{\mathrm{Mg},i}+\left(1-\phi_{i}\right)g^{(\mathrm{water})}_{\mathrm{Mg},i}\label{eq:CPF-14}
\end{align}
where $g^{(\mathrm{solid})}_{\mathrm{e},i}$, $g^{(\mathrm{solid})}_{\mathrm{Mg}++,i}$.
and $g^{(\mathrm{solid})}_{\mathrm{Mg},i}$ are the free energies
per particle, less the electrostatic contribution, for free electrons,
Mg ions, and Mg atoms respectively in the solid metal, and $g^{(\mathrm{water})}_{\mathrm{e},i}$,
$g^{(\mathrm{water})}_{\mathrm{Mg}++,i}$, and $g^{(\mathrm{water})}_{\mathrm{Mg},i}$are
the corresponding free energies per particle in water. The chemical
potentials for cell $i$ can now be found using Eq. \ref{eq:CPF-05a}.
We have 
\begin{align}
\mu_{\mathrm{Mg},i} & =g_{\mathrm{Mg},i}+n_{\mathrm{Mg},i}\frac{\partial g_{\mathrm{Mg},i}}{\partial n_{\mathrm{Mg},i}}+\Delta\mu^{(\mathrm{surface})}_{\mathrm{Mg},i}\label{eq:CPF-15}\\
\mu_{\mathrm{Mg}++,i} & =g_{\mathrm{Mg}++,i}+n_{\mathrm{Mg}++,i}\frac{\partial g_{\mathrm{Mg}++,i}}{\partial n_{\mathrm{Mg}++,i}}\nonumber \\
 & +\Delta\mu^{(\mathrm{surface})}_{\mathrm{Mg}++,i}+2ev_{i}\label{eq:CFP-15a}\\
\mu_{\mathrm{e},i} & =g_{\mathrm{e},i}+n_{\mathrm{e},i}\frac{\partial g_{\mathrm{e},1}}{\partial n_{\mathrm{e},i}}-ev_{i}\label{eq:CPF-16}
\end{align}
where 
\begin{align}
\Delta\mu^{(\mathrm{surface})}_{Mg,i} & =n_{\mathrm{Mg},i}\left(g^{(\mathrm{solid})}_{\mathrm{Mg},i}-g^{(\mathrm{water})}_{\mathrm{Mg},i}\right)\frac{\partial\phi_{i}}{\partial n_{\mathrm{Mg},i}}\nonumber \\
 & +n_{\mathrm{Mg}++,i}\left(g^{(\mathrm{solid})}_{\mathrm{Mg}++,i}-g^{(\mathrm{water})}_{\mathrm{Mg}++,i}\right)\frac{\partial\phi_{i}}{\partial n_{\mathrm{Mg},i}}\nonumber \\
 & +n_{\mathrm{e},i}\left(g^{(\mathrm{solid})}_{\mathrm{e},i}-g^{(\mathrm{water})}_{\mathrm{e},i}\right)\frac{\partial\phi_{i}}{\partial n_{\mathrm{Mg},i}}\label{eq:CPF-17}
\end{align}
and 
\begin{align}
\Delta\mu^{(\mathrm{surface})}_{Mg++,i} & =n_{\mathrm{Mg},i}\left(g^{(\mathrm{solid})}_{\mathrm{Mg},i}-g^{(\mathrm{water})}_{\mathrm{Mg},i}\right)\frac{\partial\phi_{i}}{\partial n_{\mathrm{Mg}++,i}}\nonumber \\
 & +n_{\mathrm{Mg}++,i}\left(g^{(\mathrm{solid})}_{\mathrm{Mg}++,i}-g^{(\mathrm{water})}_{\mathrm{Mg}++,i}\right)\frac{\partial\phi_{i}}{\partial n_{\mathrm{Mg}++,i}}\nonumber \\
 & +n_{\mathrm{e},i}\left(g^{(\mathrm{solid})}_{\mathrm{e},i}-g^{(\mathrm{water})}_{\mathrm{e},i}\right)\frac{\partial\phi_{i}}{\partial n_{\mathrm{Mg}++,i}}\label{eq:CPF-17a}
\end{align}
are terms that are only non-zero near the interface between the metal
and the water.

We would like a minimal model that will show the dissolution of Mg
into water as $\mathrm{Mg^{++}}$ ions, leaving electrons behind in
the metal. Thus the surface term will be neglected ($\Delta\mu^{(\mathrm{surface})}_{Mg,i}=\Delta\mu^{(\mathrm{surface})}_{Mg++,i}=0$)
as the driving term for the dissolution of the Mg into the water is
found in the bulk contributions. For bulk Mg, adding interstitials
or vacancies raises the energy. Thus we will use the following forms
\begin{align}
g^{(\mathrm{solid})}_{\mathrm{Mg},i} & =g^{(\mathrm{solid},0)}_{\mathrm{Mg}}+\chi_{\mathrm{Mg}}\frac{n_{\mathrm{Mg},i}-2n^{(\mathrm{solid},0)}_{\mathrm{Mg}}}{2n^{(\mathrm{solid},0)}_{\mathrm{Mg}}}\nonumber \\
g^{(\mathrm{solid})}_{\mathrm{Mg}++,i} & =g^{(\mathrm{solid},0)}_{\mathrm{Mg}++}+\chi_{\mathrm{Mg}++}\frac{n_{\mathrm{Mg}++,i}}{2n^{(\mathrm{solid},0)}_{\mathrm{Mg}}}\label{eq:CPF-18}
\end{align}
where $g^{(\mathrm{solid},0)}_{\mathrm{Mg}}$ is the energy per particle
in perfect bulk Mg, $n^{(\mathrm{solid},0)}_{\mathrm{Mg}}$ is the
number of Mg atoms per cell in bulk solid Mg, and $\chi_{\mathrm{Mg}}$
and $\chi_{\mathrm{Mg}++}$ are constants measuring the energy associated
with an interstitial or vacancy. In solution Mg exists as $\mathrm{Mg^{++}}$
ions with the electrostatic energy associated with it being treated
explicitly. We will thus use the forms 
\begin{align}
g^{(\mathrm{water})}_{\mathrm{Mg},i} & =g^{(\mathrm{water},0)}_{\mathrm{Mg}}\nonumber \\
g^{(\mathrm{water})}_{\mathrm{Mg}++,i} & =g^{(\mathrm{water},0)}_{\mathrm{Mg}++}\label{eq:CPF-19}
\end{align}
where $g^{(\mathrm{water},0)}_{\mathrm{Mg}}$ and $g^{(\mathrm{water},0)}_{\mathrm{Mg}++}$
are the solvation energies of an isolated Mg atom and $\mathrm{Mg}^{++}$
ion, respectively.

For the electrons we will ignore the concentration dependence in bulk
Mg, so that 
\begin{equation}
g^{(\mathrm{solid})}_{\mathrm{e},i}=g^{(\mathrm{solid},0)}_{\mathrm{e}}\label{eq:CPF-20}
\end{equation}
In solution we will assume that all electron transfer occurs into
and out of the highest energy occupied orbitals on the Mg atoms. Thus
we have a constant free energy per particle, and hence 
\begin{equation}
g^{(\mathrm{water})}_{\mathrm{e},i}=g^{(\mathrm{water},0)}_{\mathrm{e}}\label{eq:CPF-21}
\end{equation}

Substituting Eqs. \ref{eq:CPF-14}, \ref{eq:CPF-18}, \ref{eq:CPF-19},
\ref{eq:CPF-20}, and \ref{eq:CPF-21} into Eqs. \ref{eq:CPF-15},
\ref{eq:CFP-15a} and \ref{eq:CPF-16} we get 
\begin{align}
\mu_{\mathrm{Mg},i} & =\phi_{i}\left(g^{(\mathrm{solid},0)}_{\mathrm{Mg}}+\chi_{\mathrm{Mg}}\frac{n_{\mathrm{Mg},i}-n^{(\mathrm{solid},0)}_{\mathrm{Mg},i}}{n^{(\mathrm{solid},0)}_{\mathrm{Mg}}}\right)\nonumber \\
 & +\left(1-\phi_{i}\right)g^{(\mathrm{water},0)}_{\mathrm{Mg}}\label{eq:CPF-22}\\
\mu_{\mathrm{Mg}++,i} & =\phi_{i}\left(g^{(\mathrm{solid},0)}_{\mathrm{Mg}++}+\chi_{\mathrm{Mg}++}\frac{n_{\mathrm{Mg}++,i}}{n^{(\mathrm{solid},0)}_{\mathrm{Mg}}}\right)\nonumber \\
 & +\left(1-\phi_{i}\right)g^{(\mathrm{water},0)}_{\mathrm{Mg}++}+2ev_{i}\label{eq:CPF-22a}\\
\mu_{\mathrm{e},i} & =\phi_{i}g^{(\mathrm{solid},0)}_{\mathrm{e}}+\left(1-\phi_{i}\right)g^{(\mathrm{water},0)}_{\mathrm{e}}-ev_{i}\label{eq:CPF-23}
\end{align}

We now need to assign values to the parameters. As we need chemical
potential differences, but not absolute values, we set the chemical
potentials to zero in the metal. The values in the solution are then
relative to this. The electron energy change on going from $\mathrm{Mg}^{++}$
in solution to Mg metal we will take as half the bandgap of water
(9.0 eV \citep{bischoff2021band}), namely 4.5 eV. This is based on
the assumption that the energy of the solvated Mg atom's highest occupied
orbitals appear in the water band gap. The energy to move a Mg atom
from the metal into solution we construct from the following sequence
of processes: an atom is removed from the metal into vacuum (1.52
eV \citep{cox1989codata}), the atom is then ionized twice to produce
$\mathrm{Mg}^{++}$ (22.68 eV \citep{kaufman1991wavelengths}), the
electrons are returned to the metal (-7.32 eV \citep{Garron1964}),
the ion is immersed in the water (-19.0 eV \citep{Wagman1982}), and
two electrons are moved from the metal to the ion in solution (the
water band gap). The total energy change is then 6.88 eV. The energy
to transfer a $\mathrm{Mg}^{++}$ ion from the metal into solution
is then equal to the energy to transfer a Mg atom, followed by returning
two electrons to the metal, which is (6.88 - 9.0) eV = -2.12 eV. The
volume per atom for bulk Mg is 23.24 Å$^{3}$. We have used an estimate
for $\chi_{\mathrm{Mg}}$ and $\chi_{\mathrm{Mg}++}$ of 1 eV. We
have set the mesh spacing to 1 Å\ and the dielectric constant at
80 (approximately that of water) throughout the system. Similarly,
the attempt frequencies for transitions $\nu_{\alpha,i}$ are assumed
to be the same in both metal and water, and are set to $3.75\,\mathrm{fs^{-1}}$
for the electrons and $0.1\,\mathrm{fs^{-1}}$ for Mg and $\mathrm{Mg}^{++}$.

We note that this model has some similarities to that developed by
Guyer \emph{et al} for phase field modelling of electrochemical processes
\citep{Guyer2004,Guyer2004a}. We make this comparison because of
the similarity of the ambition of the two methods. The primary differences
are as follows. Here the phase field parameter is a function of the
Mg concentration rather than an independent dynamic variable: this
corresponds to a rapid transformation of phases. We ignore the interface
contributions to the energy which sets the width of the interface
region: in this case our interface thickness is set by a competition
between the electrostatic interactions (attractive between the metal
ions and the electrons, but repulsive within the electrons or within
the metal ions) and the diffusion of the particles (which tends to
broaden the interface). We also do not impose a fixed overall number
of particles in each cell (this is set by the dynamics) and do not
include the configurational entropy in our free energy (this is largely
taken into account by the dynamics).

\begin{figure}
\centering{}\includegraphics[width=1\columnwidth]{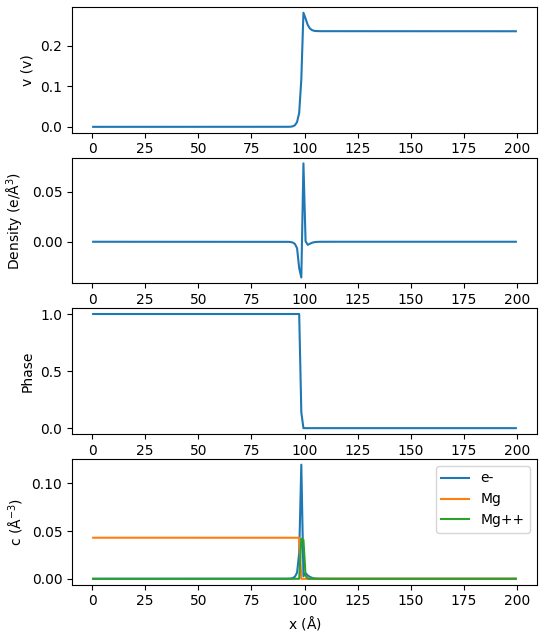}\caption{The simulation produces the expected dissolution of Mg into solution
as $\mathrm{Mg}^{++}$, with the electrons accumulating in the metal
at the interface. The potential is a result of the interfacial dipole
produced by the charge distribution. In the metal the phase has the
value 1, while in solution it has the value 0. At the interface it
can take an intermediate value.}\label{fig:dissolve}
\end{figure}

The results of a simulation are shown in Fig. \ref{fig:dissolve}.
The initial condition was bulk Mg below 100 Å and water above. Note
that maintaining stability in the simulation required the use of a
very small time step ($10^{-4}$ fs) with a forward-in-time (explicit)
Euler integrator, which yields first-order temporal convergence and
tends towards instability if the time-steps $\Delta t$ are large.

The simulation was stopped after 0.7 fs as it became unstable after
that. The results are what we would expect: Mg dissolves into the
water producing a region of $\mathrm{Mg}^{++}$ ions in the water
next to the interface, accompanied by a layer of electrons in the
metal at the interface. The two charged regions produce an electric
dipole layer, which results in a shift in the electrostatic potential
between the two sides of the interface.

We note the dissolution of Mg appears to be restricted: this might
be a result of the charge build up. For the reaction to proceed further
a mechanism to remove $\mathrm{Mg}^{++}$ from solution, and thus
removing the charge build-up at the interface, may be needed. This
could be through the formation of insoluble corrosion product ($\mathrm{Mg(OH)}_{2}$).
Alternatively, if the solution contained NaCl in high enough concentration,
Mg could combine with Cl to form neutral and soluble $\mathrm{MgCl}_{2}$.
Simulating these extra steps would involve a set of coupled reactions,
and will be left for future work.

\subsection{Diffusion in 2D and 3D}

To illustrate the extension of the KCM algorithm to higher dimensions
we will consider the problem of diffusion. The procedure is essentially
the same as in 1D, but requires extension of the discretization procedure,
and hence the writing of a new piece of software. The discretization
in time is the same as for the 1D cases considered above (first order
forward in time). It is possible to use the Runge-Kutta family or
Crank-Nicolson scheme for time-stepping. However, these higher order
methods require increased storage due to the need to include gradients
computed at differing time-steps. See the Conclusions section for
a short discussion on this point. The discretization in space for
both 2D and 3D is a natural extension if that in 1D. In 3D we have
a regular mesh with spacing of $a$ with $N_{x}\times N_{y}\times N_{z}$
cells in each dimension, where $N_{x}$, $N_{y}$, and $N_{z}$ are
positive integers. In 2D we have a regular mesh with spacing of $a$
of size $N_{x}\times N_{y}$. Rectangular prisms in 3D were not considered
as these can introduce instabilities into the solution of Poisson's
equation.

The central equations continue to be Eqs. \ref{eq:CPF-01} and \ref{eq:CPF-02}.
Note that particles can only pass between cells that share a face.
It is therefore convenient for the purpose of writing the computer
code to work with the current flowing through the boundary between
cells. The net current of particles of type $\alpha$ flowing from
cell $i$ to cell $j$ across the shared interface, $J_{\alpha,i\to j}$,
is given by 
\begin{equation}
J_{\alpha,i\to j}=n_{\alpha,i}R_{\alpha,i\to j}-n_{\alpha,j}R_{\alpha,j\to i}\label{eq:CPF3D-01}
\end{equation}
If we substitute Eq. \ref{eq:CPF-02} for $R_{\alpha,i\to j}$ into
Eq. \ref{eq:CPF3D-01} we get 
\begin{equation}
J_{\alpha,i\to j}=n_{\alpha,i}\nu_{\alpha}\exp\left(-\frac{\mu_{\alpha,j}-\mu_{\alpha,i}}{2k_{B}T}\right)-n_{\alpha,j}\nu_{\alpha}\exp\left(-\frac{\mu_{\alpha,i}-\mu_{\alpha,j}}{2k_{B}T}\right)\label{eq:CPF3D-02}
\end{equation}
Note that $J_{\alpha,j\to i}=-J_{\alpha,i\to j}$. The equation of
motion (Eq. \ref{eq:CPF-01}) then becomes 
\begin{equation}
\frac{\partial n_{\alpha,i}}{\partial t}=-\sum_{j}J_{\alpha,i\to j}\label{eq:CPF-03a}
\end{equation}
which is the continuity equation. We note in passing that a large
saving in computer time can be achieved by evaluating and storing
$\exp\left(-\frac{\mu_{\alpha,i}}{2k_{B}T}\right)$ and using it to
evaluate 
\begin{equation}
\exp\left(-\frac{\mu_{\alpha,j}-\mu_{\alpha,i}}{2k_{B}T}\right)=\frac{\exp\left(-\frac{\mu_{\alpha,j}}{2k_{B}T}\right)}{\exp\left(-\frac{\mu_{\alpha,i}}{2k_{B}T}\right)}\label{eq:CPF3D-04}
\end{equation}
In the presence of charged particles we also need to solve Poisson's
equation: see Eqs. \ref{eq:CPF-02a} and \ref{eq:CPF-02b}, though
here we consider only neutral particles.

We performed simulations of the diffusion of neutral particles in
2D to illustrate the reliability of the method. As for 1D we have
an analytic solution we can compare with. Diffusion in 2D is described
by 
\begin{equation}
\frac{\partial n_{\alpha}}{\partial t}=D\left(\frac{\partial^{2}n_{\alpha}}{\partial x^{2}}+\frac{\partial^{2}n_{\alpha}}{\partial y^{2}}\right)\label{eq:CPF3D-05}
\end{equation}
where $D$ is the diffusion coefficient. Eq. \ref{eq:CPF3D-05} is
satisfied by \citep{riley2006} 
\begin{equation}
n_{\alpha}(x,y,t)=\frac{1}{4\pi Dt}\int\exp\left(-\frac{(x-x')^{2}+(y-y')^{2}}{4Dt}\right)n_{\alpha}(x',y',0)\,\mathrm{d}x'\mathrm{d}y'\label{eq:CPF3D-06}
\end{equation}
If the initial distribution of particles has the form of a Dirac $\delta$
function centered at $x=y=0$ ($n_{\alpha}(x',y',0)=N\delta(x')\delta(y')$
where N is the total number of particles), then after a time $t$
we obtain a Gaussian distribution 
\begin{equation}
n_{\alpha}(x,y,t)=\frac{N}{4\pi Dt}\exp\left(-\frac{x^{2}+y^{2}}{4Dt}\right)\label{eq:CPF3D-07}
\end{equation}

\begin{figure}
\centering{}\includegraphics[width=0.75\columnwidth]{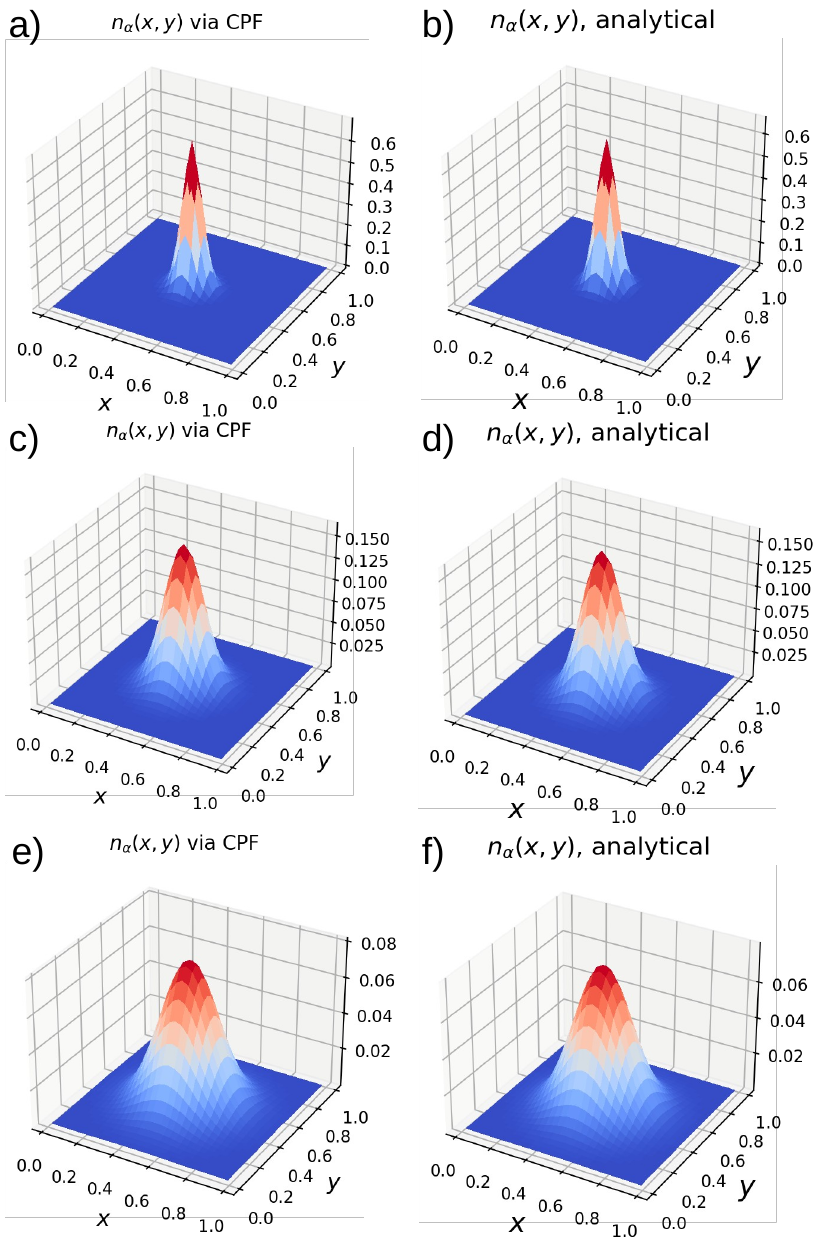} \caption{The diffusion profiles for a 2D system that starts with a very narrow
distribution in the center of the cell. The left hand figures (a,
c, e) are from the KCM simulations, while the right hand figures (b,
d, f) are from the analytic solution. The times for the figures are
0.0125 s (a and b), 0.0515 s (c and d), and 0.130 s (e and f). Note
that the labels on the axes are given as fractions of the maximum
values, taking values between 0 and 1.}
\label{fig:CPF3D-01} 
\end{figure}

Sample results for the 2D case on a square domain are shown in Fig.
\ref{fig:CPF3D-01}. The diffusion coefficient is $D=0.1\,\mathrm{m^{2}s^{-1}}$.
The agreement between the numerical and analytic solutions is very
good for $t=0.0125$ s and $t=0.0515$ s, but there is a small difference
in the peak heights for the latest time shown ($t=0.130$ s). The
difference is a result of the boundary conditions: in the analytic
calculation the boundaries are assumed to be at infinity, while for
the simulations a box of finite size is used, with the number of particles
in the box being conserved. Thus, when the tails of the diffusion
profile reach the edges of the box the particles are confined, resulting
in an increased concentration (higher peak) within the box.

\section{Conclusions}

Having described the KCM method in outline, we have shown that it
can be used to solve problems characterized by drift, diffusion, and
chemical reactions involving explicit transfer of electrons. This
provides foundations that can be built upon to simulate corrosion.
Because of the complex combinations of phases and interfaces involved,
one objective of this method is to simplify the introduction of accurate
models for the free energies of different phases, the flow of electrons,
and the description of interfaces. The focus on individual cells at,
or near, equilibrium means models of energies can be built using,
for example, Density Functional Theory calculations. This can include
phases with defects, possibly quite large in number. In addition,
by having cells as part of the definition of the model, it is hoped
that it will be easier to characterize, and hence describe, interfaces:
we have an intrinsic coordinate system relative to which models can
be constructed.

The models we have considered are of course very simple, and will
need to be extended to provide a realistic description of corrosion.
General extensions will have to include building suitable models for
2D and 3D simulations (for example, simulating roughness of interfaces,
and incorporating pitting), and improving the time integrator to ensure
simulations are stable with reasonable time steps. Beyond this, models
for corrosion product formation will need to be created, and may need
to incorporate elastic fields to describe the lattice mismatch often
found between corrosion product and the underlying metal.

A possibility for the time integrator is the following. Instead of
using a forward-in-time Euler integrator and decreasing $\Delta t$,
which adds computational expense, one could use the backward-in-time
(implicit) Euler method, which is unconditionally A-stable \citep{luochen}.
The Crank-Nicolson method combines both these integrators to yield
a second-order method. For a generic function $f$ with time derivative
$\dot{f}$, and the respective spatial and temporal discretisations
which we index by $i$ and $n$, the integration schemes are 
\begin{align}
f^{n+1}_{i} & =f^{n}_{i}+\Delta t\,{\dot{f}}^{n}_{i} & \textrm{Explicit Euler}\\
f^{n+1}_{i} & =f^{n}_{i}+\Delta t\,{\dot{f}}^{n+1}_{i} & \textrm{Implicit Euler}\\
f^{n+1}_{i} & =f^{n}_{i}+\frac{1}{2}\Delta t\left({\dot{f}}^{n+1}_{i}+{\dot{f}}^{n}_{i}\right) & \textrm{Crank-Nicolson}
\end{align}
Implementing the Crank-Nicolson scheme requires taking one explicit
Euler time-step such that $f^{1}_{i}$ and ${\dot{f}}^{0}_{i}$ are
computed, after which ${\dot{f}}^{1}_{i}$ can be computed, and the
Crank-Nicolson time-step can then be taken. It is important to note
that Crank-Nicolson comes at a higher memory expense compared to explicit
Euler. Investigating integrators will be a subject for future work.

\section*{Data availability statement}

The data that support the findings of this study are available from
the corresponding author upon reasonable request.
\begin{acknowledgments}
APH acknowledges the Thomas Young Centre under grant number TYC-101.
Discussions about the theory with Richard Fogarty are gratefully acknowledged,
as is the work of Hewen Chen and Shih-Ting Lu who used the 1D code
for their research. We also acknowledge the extensions of the computer
code that were explored by Yoonseo Lim, Mirza Sjarif, and Chenyang
Liao, though these are not used in this work. 
\end{acknowledgments}

 \nocite{*}
\bibliography{refs}

\end{document}